# Pulsars and Millisecond Pulsars I: Advancements, Open Questions and finding Gaps via statistical insights

Maria Rah[1,2],* Areg Mickaelian [1], Francesco Flammini Dotti[3], Rainer Spurzem[2,3]

[1] NAS RA V.Ambartsumian Byurakan Astrophysical Observatory (BAO), Byurakan, Armenia

[2] The Silk Road Project at the National Astronomical Observatory, Chinese Academy of Sciences, China

[3] Astronomisches Rechen-Inst., ZAH, Univ. of Heidelberg, Germany

## Abstract

Investigating the development of pulsars and millisecond pulsars (MSPs) highlights for us statistical in- sights, important research areas, and unresolved aspects to address. Through the following work, we provide a detailed demographical study of these astrophysical objects by combining data from several observational techniques (radio, X-ray, and gamma-ray) in different environments: Galactic Field (GF) and Globular Clusters (GC). Although observational studies provide direct insights into the emission properties, periodic timing (e.g. millisecond pulsars), and spatial distribution of pulsars, theoretical models are essential to interpret these findings and unravel the underlying physical processes driving their unique characteristics. We focus on the "magnetic field-spin" relationship, exploring spin-up— where accretion transfers angular momentum to the pulsar in binary systems—and spin-down, driven by magnetic dipole radiation or particle winds dissipating rotational energy. These mechanisms illuminate the intricate dynamics linking spin evo- lution to magnetic field decay. Building on these theoretical frameworks and the application of advanced numerical simulation tools such as NBODY6++GPU, CMC, and COMPASS, we provide a critical means to quantitatively test and refine our understanding of spin and magnetic field evolution in such compact objects. However, despite the advances offered by these tools, significant issues remain, particularly in in- terpreting the intricate dynamics of binary interactions involving pulsars and millisecond pulsars, as well as in accurately incorporating the physics of these compact objects into comprehensive numerical simulations. This analysis underscores the critical need for enhanced modeling frameworks and more refined observa- tional studies to address unresolved questions regarding the formation processes, evolutionary pathways, and magnetic field degradation of pulsars and MSPs. These numerical integrations of compact objects into comprehensive models of objects with a large number of stars will highlight the necessity of advancing simulation techniques. By improving these simulations, we can more accurately model the intricate physics governing pulsar behavior and ultimately resolve these outstanding challenges in astrophysics.

**Keywords:** Pulsars, Millisecond Pulsars, Statistical Insights, Evolution, Simulations, and Gaps.

## 1. Introduction

A specific type of neutron star, known as a pulsar, is characterized by its emission of periodic electro-magnetic radiation pulses, detectable across multiple wavelengths due to its intense magnetic fields and rapid rotation. Generally, pulsars typically exhibit spin periods ranging from about 0.1 to several seconds. With this unique property, normal pulsars represent a broader category of such neutron stars. However, within this category, millisecond pulsars (MSPs) stand out as a unique subclass distinguished by their extremely short spin periods, typically ranging from 1 to 10 milliseconds (Lorimer, 2008). As we mentioned, Pulsars and MSPs are unique because they are periodic, they work as accurate cosmic clocks, and they make it possible to study general relativity and dense matter physics with outstanding accuracy (Baker, 2023, Zhang et al., 2022). The stark difference in rotation speed arises from a transformative "recycling" process, where the pulsar gains angular momentum from a companion star in a binary system (Chattopadhyay, 2021). This process not only accelerates the spin of the pulsar, but also alters its magnetic field, behavior, and evolutionary pathway, differentiating MSPs from their slower-rotating counterparts. MSPs are commonly found in binary systems as neutron stars recycle material and angular momentum from their companion star (Demorest et al., 2010). The aforementioned recycling process happens quite frequently in low-mass X-ray binaries, where millisecond pulsars have periods as short as a few milliseconds (Freire et al., 2012, Lattimer & Prakash, 2007).

## 2. Historical Milestones and Future Directions for the current study

The historical evolution of pulsar research extends over nearly a century, starting with hypothesis put forward by V.Ambartsumian in 1929 concerning stars comprised of atomic nuclei (Han et al., 2021, Mickaelian, 2018), Following this claim, less than 2 years later, neutron stars were discovered by James Chadwick, and the first pulsar was identified by Jocelyn Bell Burnell and Antony Hewish in 1967 (Hewish & Burnell, 1970). Since then, advances in radio astronomy and the development of sophisticated instruments such as the Arecibo Telescope, the Parkes Multibeam Survey around 1998, and the construction of FAST, which became operational in 2017, and further accelerating discoveries have greatly expanded pulsar catalogs. See Figure 1 that shows that the rate of discovering regular pulsars surged after 2016, coinciding with advancements in survey technologies and contributions from CHIME starting in 2018 (Padmanabh et al., 2024, Sengar et al., 2023, Zhou et al., 2023).

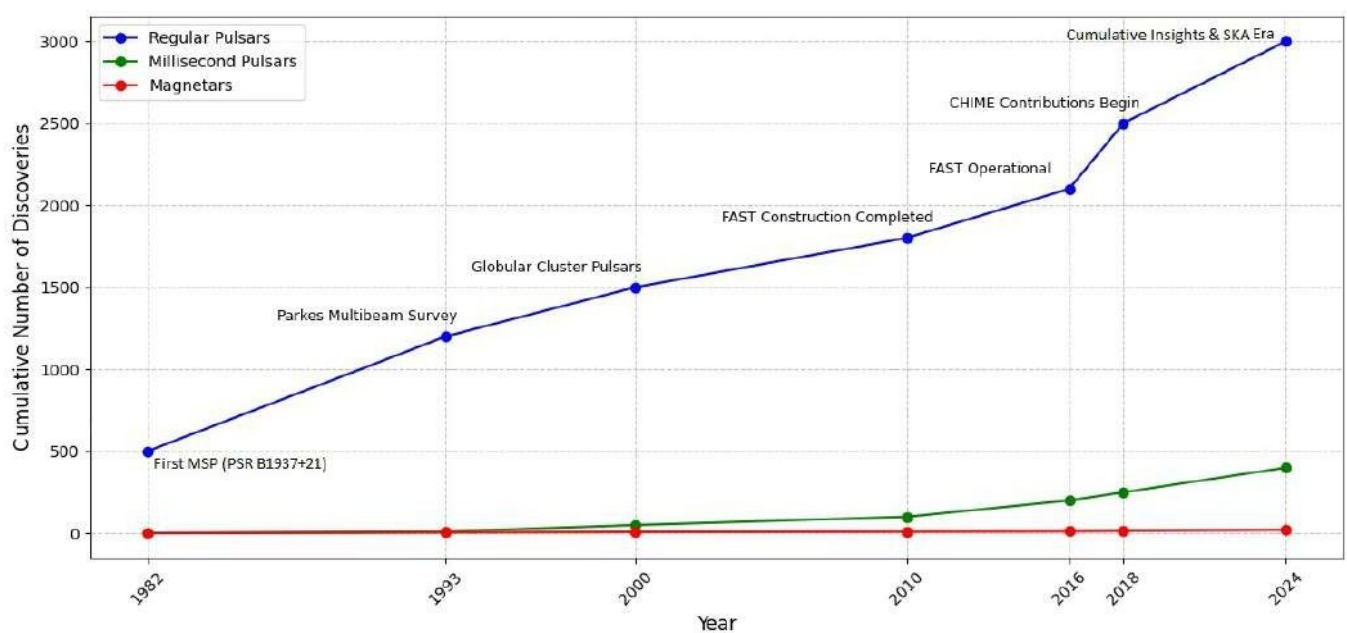

Figure 1. This figure depicts the cumulative discoveries of Regular Pulsars, Millisecond Pulsars, and Magnetars from 1982 to 2024. As is evident in the figure, the focus is on pulsars. We can find that with the growth and development of technology, which has increased the accuracy of discovery methods, the number of recorded normal pulsars is very large, and this in itself is evidence of the need to study them and their evolution pathway with greater precision and accuracy. Find more (Padmanabh et al., 2024, Sengar et al., 2023, Zhou et al., 2023)

.

Magnetars are a rare type of neutron star with extremely strong magnetic fields, billions of times stronger than typical pulsars, known for emitting intense bursts of X-rays and gamma rays (Kaspi & Beloborodov, 2017, Musolino et al., 2024). Despite steady increases in millisecond pulsars and magnetars, their numbers remain significantly lower than regular pulses, underscoring the prevalence of the latter in astronomical research. This graph effectively illustrates the impact of technological progress on the field of pulsar discovery. We now have a better understanding of these things because of later discoveries like binary pulsars that verified the predictions of gravitational waves (Falxa et al., 2023, Hulse & Taylor, 1975).

The current study serves as the first installment in a series dedicated to an statistical investigation of pulsars and millisecond pulsars as an outstanding subject of interest in contemporary astrophysics. In the following papers, which will be published in the next two issues of the current journal, we will respectively study in depth the special evolutions of pulsars and millisecond pulsars, and then, examine their evolutions during the evolutions of globular clusters by numerical simulation using NBODY6++GPU (Kamlah et al., 2022, Spurzem et al., 2023). The code will be updated for a new model of neutron star and pulsar evolution (see Sect. 7 and Ye & Fragione, 2022, Ye et al., 2019).

## 3. Statistics and observational perspectives

Pulsars and MSPs are found in different environments, including globular clusters (GC) and galactic fields (GF), which each have their own population features (Han et al., 2021). For better understanding, we refer to the cumulative number of objects found in these environments in Figure 2.

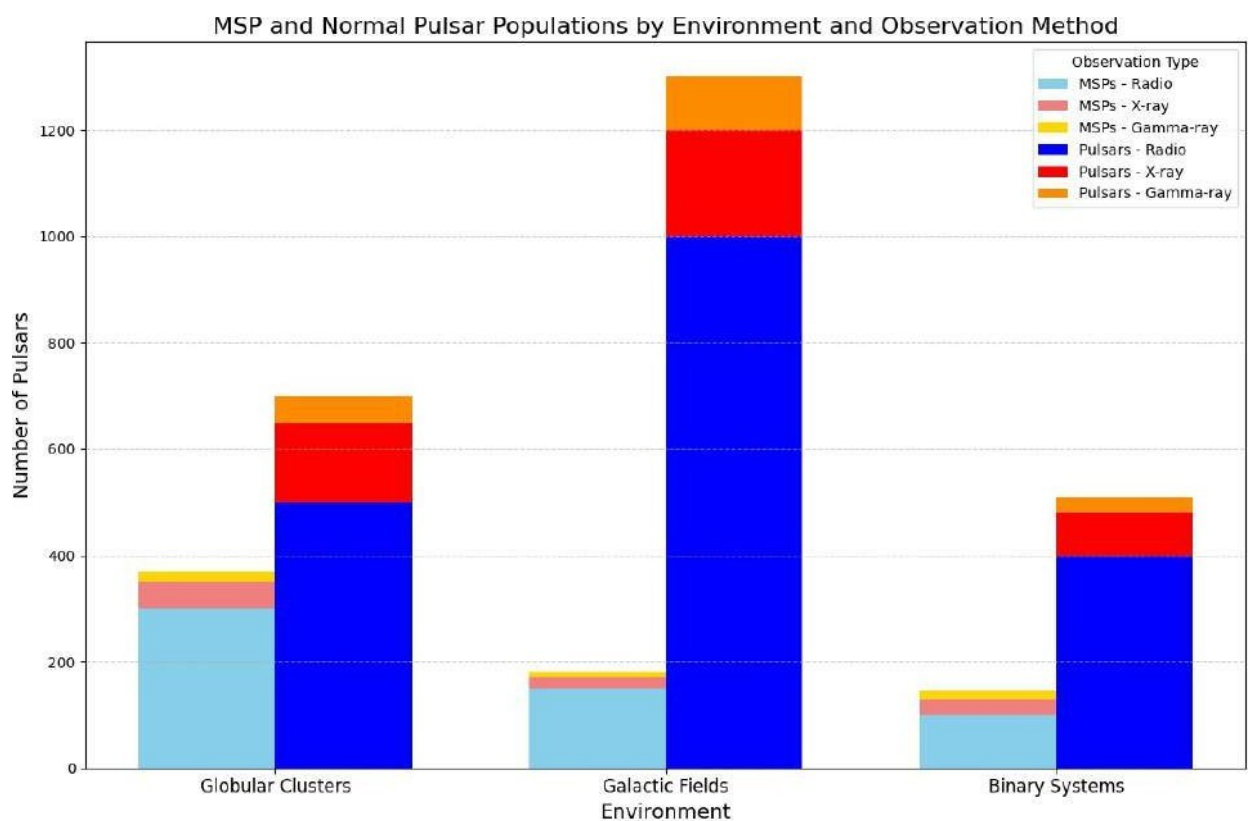

Figure 2. The chart shows the groups of millisecond pulsars (MSPs) and normal pulsars found through radio, X-ray, and gamma-ray research in three different types of astrophysical settings: globular clusters, galactic fields, and binary systems. See more in Abbott (2017), Abbott et al. (2016). The data provides a comparative analysis of how different observation approaches contribute to pulsar discovery in these specific astrophysical environments. We took the data in this figure from the ATNF Pulsar Catalog and other major pulsar research databases, with additional information from high-energy observational missions such as Fermi- LAT and Chandra.

Binary interactions that typically recycle the material of neutron stars in globular clusters often appear as MSPs, which rotate quickly and provide steady emissions that may be detected at various wavelengths. Observations such as X-ray studies from NICER and Chandra can be used to learn more about these groups. These studies show emission hotspots and accretion processes that are connected to the objects in question (Brightman et al., 2016, Deneva et al., 2019). Radio observations remain indispensable for discovering new pulsars and improving observational datasets through Pulsar Timing Arrays (PTAs) due to their unique capabilities. They excel at detecting faint pulsar signals, especially in distant globular clusters where other wavelengths fall short and enable precise timing measurements essential for gravitational wave detection via PTAs (Ferdman et al., 2010, Kramer & Champion, 2013). Similar gamma-ray observations from the Fermi Large Area Telescope (Fermi-LAT) have shown that MSPs are important high-energy sources (Abdo et al., 2013, Freire et al., 2012, Perera et al., 2019). Using more than one observational method has helped us to learn more about the populations of pulsars and how they change over time. This has shown how complex the relationship is between environmental factors and star formation (Abbott et al., 2016, Freire et al., 2017, Pan et al., 2021).

## 4. Theoretical framework

The magnetic field-spin relation (B-S) for pulsars and MSPs is influenced by their environment. Understanding these variations in the B-S relation is crucial to explaining the differences in pulsar and MSP populations in different astrophysical environments (Chen et al., 2021, Freire et al., 2012). See Figure 3 for a more detailed comparison. These diagrams are essential for understanding the effect of numerous parameters on pulsar behavior, specifically concerning "spin-up and spin-down" rates and stability.

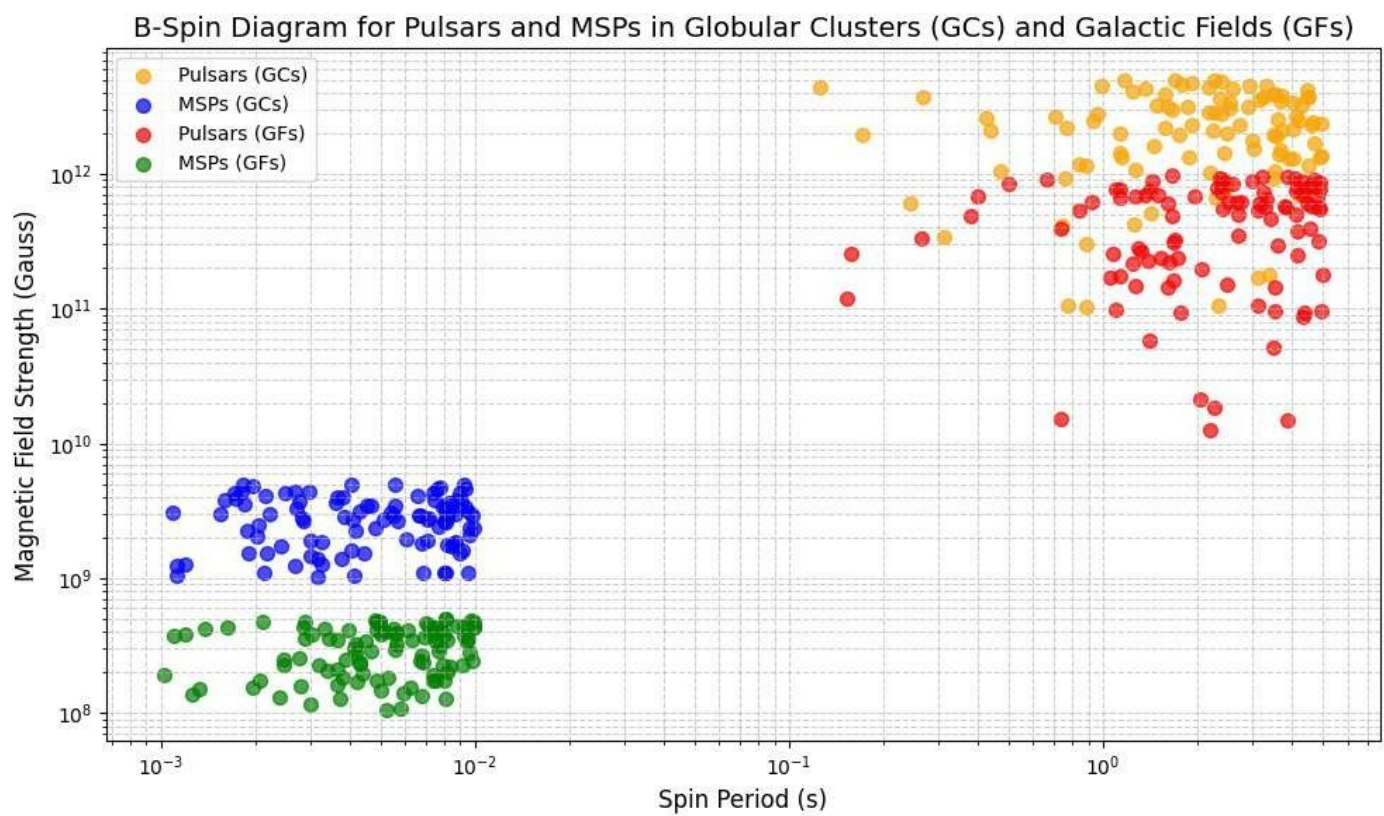

Figure 3. The B-S Diagram shows the spin periods and magnetic field of regular pulsars and millisecond pulsars (MSPs) in two different settings: Globular Clusters (GCs) and Galactic Fields (GFs). Normal pulsars are shown in orange (GCs) and red (GFs), whereas MSPs are denoted in blue (GCs) and green (GFs). The data points reflect simulated pulsar populations. See more: (Chattopadhyay, 2021, Lee et al., 2023, Ye & Fragione, 2022)

In the evolution of globular clusters, where interactions and transfer of mass in the binary are more common, the B-S relation for normal pulsars tends to show faster magnetic field decay with a quicker spin-down, while MSPs exhibit faster spin-up and weaker magnetic fields due to frequent recycling processes. In contrast, GF pulsars evolve more slowly, with a broader distribution of spin periods and magnetic field strengths, while MSPs in the GF show slower spin-up rates and weaker magnetic field decay compared to their counterparts in GCs (Lee et al., 2023). In globular clusters, millisecond pulsars (MSPs) exhibit rapid rotation rates primarily due to their interactions with companion stars, which facilitate the transfer of angular momentum during accretion events. These interactions typically occur in binary systems, where a pulsar can capture material from a nearby star, often a white dwarf or another neutron star. As the pulsar accretes this material, it gains angular momentum, leading to an increase in its rotation speed—a process known as "recycling" (Ferdman et al., 2010, Kramer & Champion, 2013). The high stellar density characteristic of globular clusters enhances the likelihood of binary interactions, allowing for frequent encounters that result in the formation of binaries. This environment significantly contributes to the rapid rotation periods observed in many MSPs, which can be less than 30 milliseconds, distinguishing them from normal pulsars (Abbott et al., 2021, Demorest et al., 2010, Freire et al., 2012). Thus, the dynamics of binary interactions within globular clusters play a crucial role in the evolution and characteristics of millisecond pulsars (Abbott et al., 2021, Kramer et al., 2021).

Besides, the magnetic field of a pulsar changes according to its environment, and thus also concerns the local stellar density and the dynamical evolution of the involved objects. This makes the evolution of these systems very complicated. The B-S diagram in Figure 3 shows how pulsars attain stable rotations via accretion processes, which in turn enhance the angular momentum transmission. The B-S relation has a feedback mechanism wherein the spin of the pulsar may affect the mass loss and development of the companion star (Chattopadhyay, 2021, Shi & Ng, 2024). Furthermore, pulsar magnetic fields may weaken with time, mainly due to accretion and interactions with the surrounding matter. This deterioration of the magnetic field can significantly affect the overall stability of the pulsar and influence its spin-down rates, as a weaker magnetic field is less effective at extracting rotational energy from the neutron star. As a result, the ability of pulsars to emit radiation efficiently is compromised, ultimately impacting their observational characteristics and lifecycle (Lattimer & Prakash, 2007, Zhang et al., 2022).

Recent models have included stellar dynamics in simulations, providing a more comprehensive understanding of MSP development in many contexts (Abbott et al., 2021, Chen et al., 2021). Researchers have found that clusters with a large number of stars, such as globular clusters, have a higher chance of having frequent stellar encounters. This means that there are more binary systems that are potential millisecond pulsars. Moreover, theoretical models indicate that binary interactions significantly influence MSP populations in both globular clusters and galactic fields(Belczynski et al., 2018, Freire et al., 2017, Perera et al., 2019).

# 5. Comparison of different numerical studies

Numerical simulations of star clusters provide a powerful tool for studying the evolution of pulsars within these environments: In these systems, the dynamics of stars is influenced solely by gravitational interactions. The use of advanced computational models, such as $\mathcal{N}$-body codes, allows researchers to simulate the complex behaviors of stellar populations in a highly populated environment, including pulsars, over time. These models are particularly advantageous for studying open and globular clusters, where many pulsars can interact with each other and with other stars in the system. $\mathcal{N}$-body simulations excel at handling large datasets, enabling the study of multiple pulsars simultaneously and observing their dynamical evolution, such as gravitational encounters or tidal disruptions, affect their spin evolution, magnetic field dynamics, and overall stability(Aarseth, 2003, Spurzem, 1999). However, these models also have limitations, as they may oversimplify the value of the parameters for certain physical processes or struggle with computational time when simulating extremely dense environments or large numbers of stars or struggle with computational time when simulating extremely dense environments or large numbers of stars. Despite these drawbacks, $\mathcal{N}$-body codes are essential for exploring intricate scenarios that involve multiple pulsars, such as the formation of millisecond pulsars (MSPs) through binary interactions or the impact of pulsar wind interactions on the surrounding stellar population. As such, these simulations offer valuable insights into pulsar evolution in star clusters, but must be continuously updated to address the complexities of real astrophysical systems once the theoretical and observational framework offer new insights. According to our statistical investigation, the NBODY6++GPU code has one of the most practical and accurate results to model dynamical interactions in dense star settings, which makes it perfect for looking at pulsars in globular clusters(Spurzem et al., 2023, Wang et al., 2015). Although NBODY6++GPU has successfully modeled the production of neutron stars, the generation of pulsars within the code is in near complete development, with ongoing improvements by its users and original creators aiming to address this aspect (Giersz & Spurzem, 1994, Spurzem, 1999). Recent improvements in GPU technology have made NBODY6++GPU run much faster and better, allowing more complex simulations to run in much less time than CPU-based models (Spurzem et al., 2023).

When comparing the three astrophysical numerical codes: NBODY6++GPU, CMC, and COMPAS, we find distinct advantages and disadvantages for each. See more comparative details at the statistical chart in Figure 4.To briefly summarize, NBODY6++GPU is good in modeling large number of self-gravitating systems (i.e., star clusters) and also adding many subroutines that reflect stellar evolution, binary gravitational interactions and 3 body interactions. It significantly outperforms others in cluster modeling, making it ideal for researchers focusing on dynamics of star clusters. You can find more about the high accuracy of NBODY6++GPU through these references(Kamlah et al., 2022, Spurzem et al., 2023, Wang et al., 2015). On the other hand, CMC, as a Monte Carlo code, takes a probabilistic approach to modeling cluster dynamics, predicting the system's evolution rather than directly calculating each timestep as done by direct N-body codes like NBODY6++GPU. While it is less computationally precise for detailed dynamical interactions, its approach allows for the modeling of large-scale systems and long-term stellar evolution in a manner that is dis- tinct from direct simulation methods. However, CMC offers a complementary approach to NBODY6++GPU, focusing on probabilistic modeling that provides insight into broader evolutionary trends, though it has po- tential for enhancement in cluster modeling and spin evolution accuracy (Ye & Fragione, 2022, Ye et al., 2019). Meanwhile, COMPAS stands out in stellar evolution and binary interactions, providing robust tools for studying binary star systems and their evolution. While it offers comprehensive features for these studies, COMPAS might not be the best choice for users who need strong capabilities in cluster modeling and magnetic field evolution, where it scores comparatively lower. Thus, each code has unique strengths and weaknesses, which makes it suitable for different aspects of astrophysical research depending on the specific needs of the study(Chattopadhyay et al., 2020).

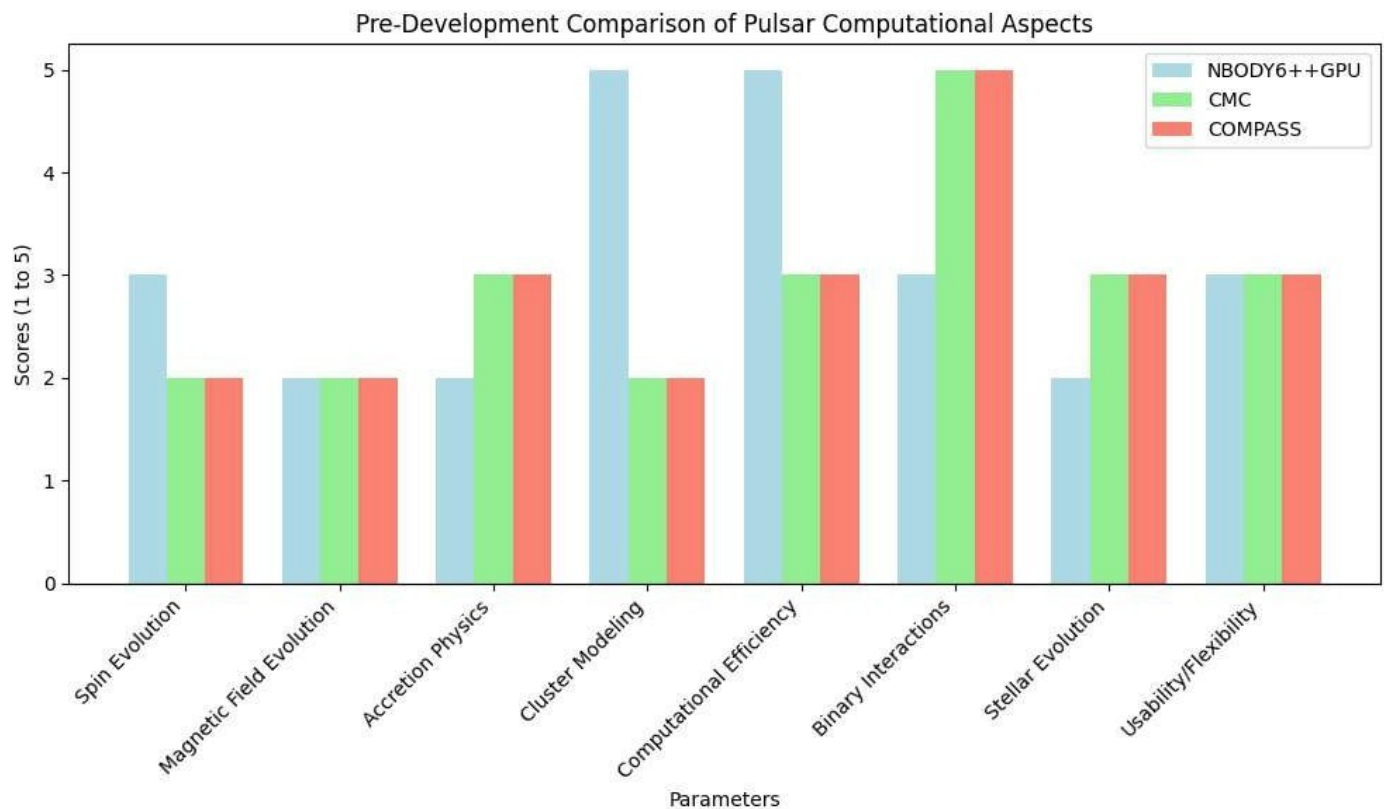

Figure 4. The graph compares three numerical studies, done through COMPAS, CMC, and NBODY6++GPU, which are derived from various development factors for pulsars such as spin evolution, magnetic field dynamics, and accretion physics. It shows that CMC and COMPAS perform similarly in most areas, especially when it comes to binary interactions and stellar evolution. However, NBODY6++GPU demonstrates notable strengths in cluster modeling and computational efficiency. See more: (Chattopadhyay et al., 2020, Wang et al., 2016, Ye et al., 2019).

NBODY6++GPU stands out among simulation software for its ability to incorporate real physical phenomena, such as tidal forces from gravitational interactions in binary systems and mass transfer between stellar companions. Its robust framework enables accurate modeling of key components of neutron star dynamics, including the spin evolution and other critical properties essential for characterizing these stars as pulsars and millisecond pulsars (MSPs). This capability makes NBODY6++GPU particularly suited for studying pulsars within the dynamic environments of star clusters, providing valuable insights into their formation and evolutionary pathways accurately. By further enhancing the code to refine the calculation of special properties of pulsars, such as spin periods and magnetic field , NBODY6++GPU can serve as a powerful tool for advancing our understanding of pulsars and MSPs in astrophysical contexts. This feature is crucial for precisely predicting the interactions that result in the creation of pulsars and MSPs in globular clusters (Wang et al., 2015), cf. (Chattopadhyay et al., 2020, Ye et al., 2019). Moreover, the incorporation of observational data into simulation frameworks has gained significant importance. By comparing simulation results with observational data, like the Fermi Large Area Telescope (Fermi-LAT) and the Chandra X-ray Observatory, researchers can test their models and learn more about pulsar populations (Freire et al., 2017, Perera et al., 2019). The synergy between simulations and observations is vital for resolving outstanding problems about MSP creation processes and their evolutionary paths (Abbott(LIGO) et al., 2017).

## 6. Outstanding research questions

Future investigations will try to answer the following unanswered research questions:

1. Formation Mechanisms: What specific processes lead to the formation of MSPs within different environments like GCs and GFs? (Chattopadhyay, 2021, Chattopadhyay et al., 2020, Lee et al., 2023, Ye & Fragione, 2022, Ye et al., 2019).

2. Binary Interactions: How do various binary configurations—such as structural properties (e.g., spin and magnetic field of neutron stars) and orbital parameters (e.g., semi-major axis and eccentricity)—influence the spin-up processes of neutron stars (Arca Sedda et al., 2023, Galloway et al., 2024).

3.Magnetic Field Evolution: What mechanisms govern the decay or enhancement of magnetic fields in pulsars over timescales ranging from millions to billions of years, and how do these processes differ between pulsars in globular clusters, influenced by dense stellar interactions, and those in the relatively isolated environments of the Galactic field (Igoshev et al., 2021, Wang et al., 2016, Zhang et al., 2022).

4. Gravitational Wave Detection: Pulsar timing arrays (PTAs) have revolutionized astrophysics by enabling the detection of gravitational waves through the precise monitoring of millisecond pulsars. These arrays are particularly sensitive to low-frequency gravitational waves from cosmic phenomena such as supermassive black hole mergers. Building on this capability, how can we further optimize PTAs to enhance their sensitivity and detect gravitational waves from even more distant or subtle cosmic events?( Abbott(LIGO) et al., 2017, Lee et al., 2023).

5. Probing the Equation of State with Pulsars: How can pulsars, under the extreme conditions -ultra high densities, intense gravitational fields, and rapid rotation- provide critical insights into the equation of state for nuclear matter, particularly in the dense cores of neutron stars?(Lattimer, 2021, Zhang et al., 2022).

6. Role of Pulsars in Cosmic Evolution: How do pulsars contribute to cosmic evolution, and in what ways do their roles and influences on stellar dynamics differ between the dense, interaction-rich environments of globular clusters and the relatively isolated conditions of the Galactic field?(Chattopadhyay, 2021, Chattopadhyay et al., 2020, Ye et al., 2019).

These questions emphasize that modeling the evolutionary processes of pulsars and MSPs requires advancements in theoretical frameworks and simulations.

## 7. Discussion and future research

The relevance of observational missions using instruments like FAST, CHIME, and NICER is highlighted by recent developments in the detection and characterization of pulsars and millisecond pulsars (MSPs). Because of these efforts, pulsar catalogs have been greatly expanded, revealing exotic systems like transitional MSPs[1] and black widow pulsars[2]. Pulsar Timing Arrays (PTAs) have attained nanosecond accuracy, facilitat- ing progress in gravitational wave astronomy ( Perera et al., 2019). These statistical findings have improved our understanding of pulsar distributions and evolutionary pathways, particularly within the

---

[1]Millisecond pulsars that alternate between accretion-powered X-ray states and rotation-powered radio states, showcasing their evolving nature in binary systems.

[2]These are millisecond pulsars in binary systems that emit intense radiation, often ablating their low-mass companion stars over time.

Galactic Field (GF) and contrasting globular clusters (GCs). Theoretical advancements in pulsar evolution, especially regarding spin and magnetic field dynamics, have been clarified using advanced numerical simulations. During the last years, it was possible to numerically simulate their mechanisms, like accretion, tidal spin-up, and gravitational interactions in dense settings using numerical tools such as NBODY6++GPU, CMC, and COMPAS. These codes have shown how stellar dynamics, pulsar recycling in globular clusters, and the galactic field interact with each other (Liu et al., 2021, Zhang et al., 2022). NBODY6++GPU has effectively simulated the formation of neutron stars, yet the inclusion of pulsar generation within the code represents a promising area for further exploration which is soon going to be explored. The Nbody6++GPU team is working to enhance its capabilities to address this exciting challenge. Expanding this system to incorporate features tailored to pulsars will be crucial for advancing our understanding of these fascinating objects and their diverse types. Very accurate modeling of pulsar populations may help researchers resolve inconsistencies between theoretical predictions and observed data. Our statistical review of the recent progress in pulsars and MSPs revealed that these developments are crucial for accurately analyzing pulsar dynamics and interactions within astrophysical contexts (Chattopadhyay, 2021, Chattopadhyay et al., 2020, Ye & Fra- gione, 2022, Ye et al., 2019). In conclusion, addressing these challenges through improved numerical models will deepen our understanding of pulsars and MSP formation and enable more rigorous testing of advanced astrophysical theories. The ongoing development of NBODY6++GPU presents an exciting opportunity to explore the complex mechanisms underlying pulsar evolution with greater precision.

## Acknowledgments

FFD and RS acknowledge support by the German Science Foundation (DFG), priority program SPP 1992 "Exploring the diversity of extrasolar planets" (project Sp 345/22-1 and central visitor program), and by DFG project Sp 345/24-1.